\documentclass[twocolumn, secnumarabic, amssymb, amsmath,nobibnotes, aps, prl,nofootinbib]{revtex4}

\usepackage{graphicx}

\date{October 22, 2011}

\newcommand{\twothirds}{\mbox{\small{$\frac{2}{3}$}}}

\newcommand{\fourthirds}{\mbox{\small{$\frac{4}{3}$}}}
\begin{document}
\title{what's wrong with black hole thermodynamics?}
\author{B. H. Lavenda}
\email{info@bernardhlavenda.com}
\homepage{www.bernardhlavenda.com}
\affiliation{Universit$\grave{a}$ degli Studi, Camerino 62032 (MC) Italy}
\begin{abstract}
 Not only is the Bekenstein expression for the entropy of a black hole a convex function of the energy, rather than being a concave function as it must be, it predicts a final equilibrium temperature given by the harmonic mean. This violates the third law, and the principle of maximum work. The property that means are monotonically increasing functions of their argument underscores the error of transferring from temperature means to means in the internal energy when the energy is not a monotonically increasing function of temperature. Whereas the former leads to an increase in entropy, the latter lead to a decrease in entropy thereby violating the second law. The internal energy cannot increase at a slower rate than  the temperature itself.
\end{abstract}
\maketitle

The fallaciousness of the Bekenstein expression~\cite{JB} for the entropy of a black hole is made apparent by the following derivation. 

Consider the relativistic thermal wavelength,
\begin{equation}
\lambda=\frac{\hbar c}{kT}. \label{eq:thermal}
\end{equation}
The only length that can be associated with a black hole is the Schwarzschild radius, so we set the thermal wavelength equal to it to obtain~\cite{Sexl}
\begin{equation}
T=\frac{E^2_{\mbox{\tiny Pl}}}{2kE}, \label{eq:Hawking-T}
\end{equation}
upon rearrangement, where $E_{\mbox{\tiny Pl}}=\surd\left(\hbar c^5/G\right)$ is the Planck energy, and where  the internal energy, $E$, has been confused with the rest energy, $mc^2$. 

Hawking~\cite{SH} claims that tiny black holes will radiate energy away at temperature \eqref{eq:Hawking-T} leading to complete evaporation in  an explosion that would result in an intense burst of photons in the x-ray region. 

Black bodies do not evaporate; in fact, Stefan's law is derived directly  from the Clapeyron equation, 
\begin{equation}
\frac{dp}{dT}=\frac{h}{T},\label{eq:Clap}
\end{equation}
where $h$ is the enthalpy density, or the latent heat in this context. 

The Clapeyron equation, \eqref{eq:Clap}, characterizes a first-order phase equilibrium. The analogy between thermal radiation and a solid-gas equilibrium, with the gas consisting of photons and the solid being the cavity walls which are both source and sink for photons, has long been known.~\cite{JV} 

If the cavity walls expand isothermally, photons will be injected into the cavity bringing with them a quantity of heat $Q=\fourthirds\sigma T^4\Delta V$, where $\sigma$ is the radiation constant and $\Delta V$ is the change in the volume of the cavity. The ratio, $Q/\Delta V=h$, is the latent heat of sublimation. The volume can be varied independently so as to keep the vapor pressure constant. 

From the phase rule, $f=r-m+2$, the number $f$ of intensive parameters that can be varied independently is $1$, if the number of components $r=1$, and the number of phases $m=2$. This is, indeed, verified: the chemical potential vanishes in the presence of an unconserved number of particles, and the pressure, $p=p(T)$, is the single intensive variable that can be varied independently. Under isobaric conditions there can be no change in the temperature, and, hence, $C_p$ cannot be defined for  black body radiation.~\cite{RK} \emph{Hence, black body radiation cannot be used as a mechanism of black hole evaporation\/}.

Appealing to the second law,
\begin{equation}
\frac{dS}{dE}=\frac{1}{T}=k\frac{2E}{E^2_{\mbox{\tiny Pl}}}, \label{eq:II}
\end{equation}
we obtain 
\begin{equation}
S=k\frac{E^2}{E_{\mbox{\tiny Pl}}}+\mbox{const.}, \label{eq:Bekenstein}
\end{equation}
upon integration. Expression  \eqref{eq:Bekenstein} is to be considered as the fundamental 
relation for a black hole, and, since it does not contain any mechanical variables, no work processes are contemplated. 

Differentiating \eqref{eq:II} a second time gives
\[
\frac{d^2S}{dE^2}=-\frac{1}{T^2}\frac{dT}{dE}=-\frac{1}{T^2C}=\frac{2k}{E^2_{\mbox{\tiny Pl}}}, \]
so that the heat capacity, $C$ is
\begin{equation}
C=-\frac{E^2_{\mbox{\tiny Pl}}}{2kT^2}. \label{eq:C}
\end{equation}
The fact that $C<0$, we are told, should not dissuade from considering \eqref{eq:Bekenstein} as the true entropy of a black hole because the heat capacities of many stars are negative. 

However, those heat capacities have nothing to do with the second law. They refer to \emph{polytropes\/}, which are defined by an equation of state of the form:
\[ pV^n=\mbox{const}.\]
If during the process of stirring, a quantity of heat, $dQ$, is supplied then it will be proportional to the change in temperature, $dt$, viz.,
\begin{equation}
dQ=c dt, \label{eq:C-bis}
\end{equation}
where $t$ is the empirical temperature. Expression \eqref{eq:C-bis} is the definition of a polytropic change.~\cite{SC} If the constant, $c$, is zero, the change is adiabatic, while if it is infinite, the change is isothermal. It was this  \lq heat capacity\rq\ that was thought can become negative.~\cite{PTL} Even if this were true, it has nothing to do with the second law.

Consider an ideal gas: $pV/t=A:=C_p-C_V$, where $C_p$ and $C_V$ are the heat capacities at constant pressure and volume, respectively. The doctrine of latent and specific heats demands~\cite{L09}
\begin{equation}
dQ=C_Vdt+L_VdV \label{eq:doc}
\end{equation}
if the pair of independent variables are $(t,V)$, and $L_V$ is the latent heat. Introducing \eqref{eq:C-bis} and dividing through by $t$ results in
\begin{equation}
\frac{dt}{t}+(n-1)\frac{dV}{V}=0, \label{eq:adiabat}
\end{equation}
since $L_V=p$ for an ideal gas, and where the polytropic index $n$ is defined as~\cite[p. 41]{SC}:
\begin{equation}
n=\frac{C_p-c}{C_V-c}. \label{eq:n}
\end{equation}
Integrating \eqref{eq:adiabat} gives
\begin{equation}
V^{n-1}t=\mbox{const}. \label{eq:z}
\end{equation}

Thus,
\begin{align}
dQ &=C_Vdt+pdV=C_Vdt-\frac{pV}{t(n-1)}dt\nonumber\\
&=\left[C_V-\frac{A}{n-1}\right]dt \label{eq:doc-bis}
\end{align}
so that
\begin{equation}
c=\frac{C_V(n-1)-A}{n-1} \label{eq:c-bis}
\end{equation}
can become negative in the interval~\cite{PTL}
\begin{equation}
1<n<\gamma,\label{eq:ineq}
\end{equation}
where $\gamma:=C_p/C_V$.

However, it is important to bear in mind:
\begin{itemize}
\item temperature and volume are no longer independent quantities, but, rather are related to the adiabatic relation, \eqref{eq:z}, and
\item this has nothing to do with the second law.
\end{itemize}
It is the concavity of the entropy which is the essence of the second law~\cite{BL95}, and which the Bekenstein entropy, \eqref{eq:Bekenstein},  violates. We are now going to show that for an ideal, monatomic, gas the heat capacity \eqref{eq:c-bis} always vanishes because the second inequality in \eqref{eq:ineq} is actually an equality. 

The polytropic index, $n$, is related to the Gr\"{u}neisen parameter, $s$, defined by the equation of state~\cite[p. 40]{L09},
\begin{equation}
s=pV/E. \label{eq:Grun}
\end{equation}
For the ideal gas under consideration, 
\begin{equation}
s=\gamma-1=\twothirds. \label{eq:s}
\end{equation}
 The relation between $s$ and $n$ follows from the fact that $V^s$ is the integrating factor for~\cite[p. 42]{L09}
\[dQ=dE+pdV=dE+s\frac{E}{V}dV.\]
Upon multiplication on both sides by $V^s$ there results the perfect differential,
\[V^sdQ=d\left(EV^s\right),\]
implying that $tV^s=\mbox{const}.$ is the adiabatic condition for an ideal gas. Comparing this with \eqref{eq:z} we  conclude that $s=n-1$, and $n>1$. According to \eqref{eq:s}, $n=\gamma$, and from its definition, \eqref{eq:n}, we conclude that $c$ vanishes. \emph{Thus, for an ideal gas, $c$ can never be negative\/}.

We now divide the black hole up into  number of cells,  with masses $m_j$, and with  specific heats, $c(T_j)$.  Initially, they are separated by adiabatic impermeable walls which are subsequently replaced by diathermal walls. We want to compare greater and less constrained states of equilibrium, and, thereby, determine the final common temperature.~\cite{EC} 

The heat capacity \eqref{eq:C}, derived from the Bekenstein entropy, \eqref{eq:Bekenstein}, would lead to a conservation of energy,
\[
\Delta E=-c\sum_jm_j\int_{T_{j}}^{T_{f}}\frac{dT_j}{T_j^2}=c\left\{\frac{M}{T_f}-\sum_j\frac{m_j}{T_j}\right\}=0,\]
from which we would deduce the final equilibrium temperature as the weighted \emph{harmonic\/} mean temperature,
\begin{equation}
\frac{1}{T_f}=\frac{1}{M}\sum_j\frac{m_j}{T_j}, \label{eq:harmonic}
\end{equation}
where
\[M=\sum_jm_j,\]
is the total mass. However, \eqref{eq:harmonic} would violate the third law of thermodynamics,  which, in its strong form reads
\begin{equation}
\lim_{T\rightarrow0} S=0. \label{eq:III}
\end{equation}
\emph{It is, therefore, necessary to assume that the $T_j$ are nonnegative which implies that all means of order less than zero be taken as zero\/}.~\cite{EB}

For simplicity consider two subsystems. Then multiplying \eqref{eq:harmonic} through by $cM$ gives
\begin{equation}
E_f=\frac{cM}{T_f}=\left(E_1+E_2\right)=\frac{cm_1}{T_1}+\frac{cm_2}{T_2}, \label{eq:arithmetic}
\end{equation}
which says simply that energy is conserved.
 Likewise, the second law reads
\[
\Delta S=c\left\{\frac{M}{T_f^2}-\left(\frac{m_1}{T_1^2}+\frac{m_2}{T_2^2}\right)\right\}\ge0,\]
which, in terms of the energies, implies
\begin{equation}
\frac{E_f^2}{M}\ge\frac{E_1^2}{m_1}+\frac{E_2^2}{m_2}.\label{eq:wrong}
\end{equation}
Again, for simplicity, suppose  $m_1=m_2=m$ so that $M=2m$, and \eqref{eq:wrong} reduces to
\[
\left(E_1+E_2\right)^2\ge2\left( E_1^2+E_2^2\right). \]
This simplifies to
\[2E_1E_2\ge E_1^2+E_2^2,\]
which is blatantly wrong. 

If we call $\mathfrak{M}_p$, the mean of order $p$, then because it is a monotonic increasing function of $p$~\cite{GH},
\begin{equation}
\mathfrak{M}_{p+1}\ge\mathfrak{M}_p,\qquad \forall -\infty<p<+\infty. \label{eq:MM}
\end{equation}
But because $E$ is \emph{not\/} a monotonically increasing function of $T$, we cannot use \eqref{eq:MM} to guarantee that the means of the energy will behave in the same way as the temperature. This is an internal consistency condition of thermodynamics meaning that it outlaws means of order $p<0$.

Thus, it is not like what has been claimed: \lq\lq heat capacities proportional to negative powers of $T$ can well be considered except at very low $T$.\rq\rq~\cite{SS} Thermodynamics is self-consistent, and violating any one of its laws leads to internal contradictions in the others.

If \eqref{eq:harmonic} were a viable, common final temperature, it would violate the principle of maximum work which identifies the geometric mean as the final common temperature. This occurs for the ideal gas in which the transition from initial to final states is \emph{reversible\/},~\cite{Callen}
\begin{align*}
\Delta S&=c\sum_jm_j\int_{T_{j}}^{T_{f}}\frac{dT_j}{T_j}\\
&=c\left\{M\ln T_f-\sum_jm_j\ln T_j\right\}=0, 
\end{align*}
leading to the weighted geometric mean temperature,
\begin{equation}
T_f=\left(\prod_jT_j^{m_j}\right)^{1/M}, \label{eq:geometric}
\end{equation}
as the final mean temperature. The maximum work would then follow from the first law,
\begin{align}
W&=-\Delta E=-c\sum_jm_j\int_{T_{j}}^{T_{f}}dT_j\nonumber\\
&=c\left\{\sum_jm_jT_j-\left(\prod_jT_j^{m_j}\right)^{1/M}\right\}>0, \label{eq:arith-geo}
\end{align}
where the inequality follows directly from the arithmetic-geometric mean inequality. 

\emph{This shows that the internal energy cannot increase more slowly than the temperature itself, and, consequently, rules out thermal equations of state of the form \eqref{eq:Hawking-T}\/}. This is entirely reasonable since $kT$ is the average kinetic energy with which the particles in a gas move about. The internal energy may increase faster than the temperature due to creation and annihilation of particles, such as in a photon gas, but it cannot increase at a slower rate.

\end{document}